\documentclass[12pt,a4paper]{article}

\usepackage{graphics}
\usepackage{graphicx}
\usepackage{xspace}

\usepackage{amsmath,amssymb}

\usepackage[dvips]{color}

\usepackage{wrapfig}
\def\lb{\linebreak[4]}
\newcommand{\be}{\begin{equation}}
\newcommand{\ee}{\end{equation}}
\newcommand{\bea}{\begin{eqnarray}}
\newcommand{\eea}{\end{eqnarray}}
\newcommand{\bes}{\begin{subequations}}
\newcommand{\ees}{\end{subequations}}
\newcommand{\bear}{\begin{equation}\begin{array}}
\newcommand{\eear}[1]{\end{array}\label{#1}\end{equation}}

\def\ba{$$\begin{array}}
 \def\ea{\end{array}$$}
\newcommand{\beg}{\begin{equation}\begin{gathered}}
\newcommand{\eeg}{\end{gathered}\end{equation}}
\newcommand{\beal}{\begin{equation}\begin{aligned}}
\newcommand{\eeal}{\end{aligned}\end{equation}}
\newcommand{\begg}{\begin{gather*}}
\newcommand{\eegg}{\end{gather*}}

\newcommand{\fr}[2]{\dfrac{{ #1}}{{ #2}}}

\newcommand{\la}{\langle}
\newcommand{\ra}{\rangle}
\newcommand{\fn}[1]{\footnote{{#1}}}
\newcommand{\bu}{$\bullet$\ }
\newcommand{\tra}{$\triangledown$ }

\renewcommand{\le}{\leqslant}
\renewcommand{\ge}{\geqslant}

\newcommand{\epe}{\mbox{$e^+e^-\,$}}
\newcommand{\ggam}{\mbox{$\gamma\gamma\,$}}
\newcommand{\egam}{\mbox{$e\gamma \,$}}

\def\cl{\centerline}

\newenvironment{Itemize}{\begin{list}{$\bullet$}%
{\setlength{\topsep}{0.2mm}\setlength{\partopsep}{0.2mm}%
\setlength{\itemsep}{0.2mm}\setlength{\parsep}{0.2mm}\setlength{\leftmargin}{4mm}}}%
{\end{list}}
\newcounter{enumct}
\newenvironment{Enumerate}{\begin{list}{\arabic{enumct}.}%
{\usecounter{enumct}\setlength{\topsep}{0.2mm}%
\setlength{\partopsep}{0.2mm}\setlength{\itemsep}{0.2mm}%
\setlength{\parsep}{0.2mm}\setlength{\leftmargin}{4mm}}}
{\end{list}}

\begin{document}

\title{High Energy Photon Collider}
\author{I. F. Ginzburg and G. L. Kotkin\\
 Sobolev Institute of Mathematics,  Novosibirsk, 630090,  Russia\\
 Novosibirsk State University, Novosibirsk, 630090,  Russia}

\date{}

\maketitle
\abstract{
We discuss a high-energy photon linear collider (HE PLC) based on the \epe \  linear collider with cms electron energy $2E = 1 \div 2$~TeV (JLC, CLIC,...).
This energy region was previously considered hopeless for experiment.
On the contrary, the present study leads to a rather optimistic conclusions.
We compare properties of HE PLC with those  of the usually discussed {\it standard PLC} with $ E\approx 250$~GeV.  We show that at the optimal choice of laser the high-energy \ggam luminosity integral  is about 1/5, and the maximum luminosity is about 1/4 from similar values  for the standard PLC.
For this choice, the laser flash energy and laser-optical system should be approximately the same as those prepared for the standard PLC.
 The photon spectrum of HE PLC is much more monochromatic than  that in the standard PLC, it is concentrated near the high-energy limit with an energy spread of about 5\%. It will be well separated from the low energy part.

}
\maketitle

\section{Introduction}

\paragraph{Two photon  processes -- virtual vs real photons.}
A process belonging to the class of what is now called two-photon was first considered in 1934.  This was the production of \epe \ pairs in the collision of ultrarelativistic charged particles, $A_1A_2 \to A_1A_2+X$ with $X= \epe$ \cite{LL}. Next 35 years different authors considered similar processes with  $X=\mu^+\mu^-$, $X=\pi^+\pi^-$ (point-like), $X=\pi^0$ (see references in review \cite{BGMS}).
In 1970, it was shown that observing the processes $A_1A_2 \ to A_1A_2 + X$ at colliders (primarily at \epe colliders) will allow us to study the processes $\ggam \to X$ with two quasi-real photons at very high energies \cite{BBG} (and a little later \cite{BKT}).

The general description of such processes given in the review \cite{BGMS} is relevant till now.
The collision of particles $A_i$ with mass $M_i$,  electric charge $Z_ie$ and energy $E_i$ produces  two
{\it virtual} (quasi-real) photons with energies $\omega_i$.  Their fluxes (per initial particle $A_i$) are
$$
f(\omega_i)d\omega_i\approx\fr{Z_i\alpha}{\pi}\,g_i(\omega_i/E_i)\,\fr{d\omega_i}{\omega_i}\,L_i\;\theta(L_i)\,,\quad
L_i\approx \ln \left(\lambda_q^2(E_i/M_i)^2/\omega_i^2\right)\,,
$$
where the form of the function $g_i (x) \le 1$ and the parameter $ \lambda_q $ depend on the type of collided particle $ A_i $ and on the properties of the system $X$. Thus, the experiments with quasi-real photons became a natural part of the collider experiments (see, for example, \cite {PHOTON17}). They made important additions to our knowledge of the resonances and details of hadron physics.  However, they are not competitive with the studies of New Physics on \epe  and $pp $ colliders. Indeed,  in the high-energy part the luminosity of virtual photon collisions is 3-5 orders of magnitude lower than that  of the base \epe or $pp$ collider.   For a collision of heavy nuclei (RHIC, LHC), the effective energy spectrum of virtual photons is limited from above and difficulties with the signature of \ggam events are added. Therefore, studying the effects of New Physics is very difficult or impossible in  collision of virtual photons at all these colliders.

Quite different approach allowing to obtain \ggam collisions with the {\it real} photons and to use them for study the phenomena of New Physics was developed in  1981 \cite{GKST} when discussing the potential of \epe linear colliders (with electron energies $E$). In these colliders, each electron bunch is used only once. Therefore, one can try to convert a significant fraction of incident electrons into photons with energy $\omega \sim E$ so that the resulting \ggam collisions  compete with the main \epe collision in both the collision energy and luminosity.

\paragraph{Photon Colliders with real photons -- PLC.}

This option -- Photon Linear Collider (PLC) was proposed by our group  \cite{GKST} as  a specific mode of future \epe\ linear collider (LC)  \cite{TESLA}-\cite{CLICmod}. PLC will be useful both for clarifying the results obtained at the hadron collider (LHC type) or \epe  colliders -- linear (ILC, CLIC,...) or circular (FCC-ee, CEPC,...) -- and for solving New Physics problems that are inaccessible or very difficult to study at these colliders

\begin{figure}[hbt]\centering{
\includegraphics[height=0.22\textheight,
width=0.8\textwidth]{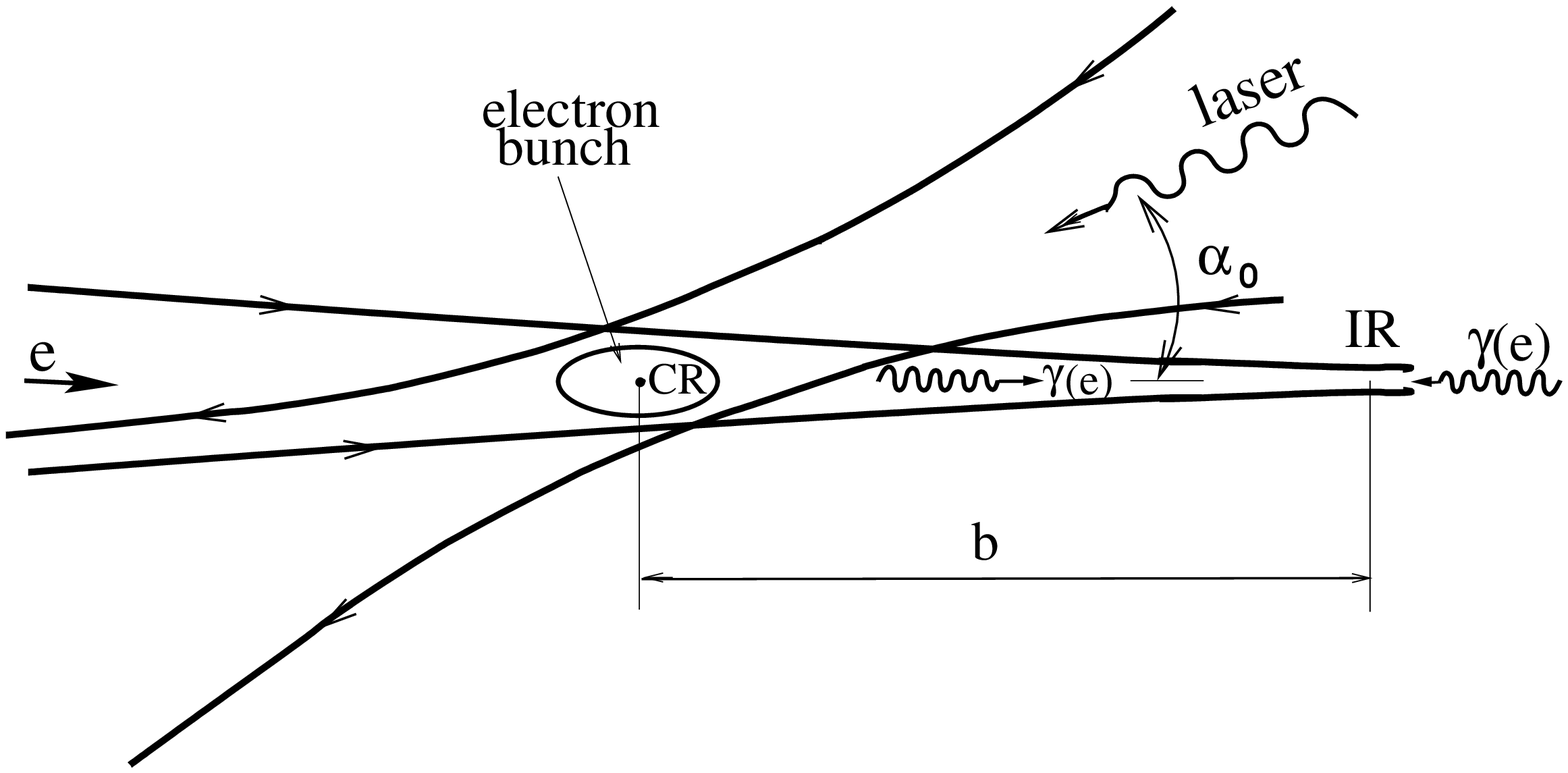}\vspace{-5mm}
\caption{\it Scheme of PLC } \label{fig:basschem}}
\end{figure}

The well-known scheme of \  PLC
is shown in fig.~\ref{fig:basschem}. At conversion  region
 $CR$ preceding the  interaction  region $IR$,
 electron ($ e^-$ or $ e^+$) beam of the basic LC
meets the photon flash from the powerful laser.
The Compton backscattering of laser photons on electrons from LC produces high
energy photons. With   the suitable choice of laser one can
obtain photon  beams with the photon energies
close to that of the initial electron. These photons
are focused in the interaction region  at the approximately the same spot, as it
was expected for electrons without laser conversion.  As a result,
the \ggam \ or \egam \ collisions under interest occur in the interaction region\fn{The Coulomb repulsion in the IR reduces the luminosity of $e^-e^-$ collisions. Besides, in the SM these collisions cannot produce interesting final states with large effective mass}.
The ratio of number of high energy photons to that of electrons --
the conversion coefficient $k\sim 1$, for the standard PLC $k\approx
1-1/e\approx 0.63$ typically.  Numerous studies of PLC (see, for example,
\cite{TESLA}, \cite{Tel16}) are developing many new technical details\fn{ Example: optimization of the beam crossing angle  for \epe and \ggam  collisions  at the LC \cite{Tel18}.}, but
all of them keep the initial scheme fig.~\ref{fig:basschem}.

Main properties of the basic Compton process are determined by the  parameter
 \be
 x=4E\omega_0/m_e^2\,,
\label{xdef}
\ee
where $E$ is the electron beam energy and
$\omega_0$ --  the laser photon energy. (To simplify
text, we  set $\alpha_0$ in Fig.~\ref{fig:basschem} to be zero.)

The most suitable modern lasers with
neodymium glass or garnet allow to realize such scheme in its
pure form only for   the electron beam energy $E\le 270$~GeV
(first stage of ILC). For $x>2(1+\sqrt{2})\approx 4.8$ (at
$E>270$~GeV using the same laser) some of produced high energy
photons   die out, producing  \epe pairs in
the collisions with laser photons from the tail of laser flash.
This fact was treated as limiting for realization of PLC based on
LC with higher electron energy \cite{GKST}).  To overcome
this difficulty,   two  alternatives were discussed:

\tra to use a new laser with lower photon energy;

\tra to use existing lasers and accept a reduction in \ggam \ lu\-mi\-nosity.

 \tra For the first alternative, the results of \cite{GKST} are applicable.
Only for each new energy of electrons a new laser is needed, and  such lasers with the necessary parameters do not exist.

\tra We consider the second alternative, following the idea that it is important to generate clean photon collisions even at the expense of some reduction in luminosity.\\ This opportunity was mention in
papers \cite{Tel2001,GKPho9} but the detailed
analysis of processes appeared at $x>4.8$, optimization of condition for $e\to \gamma$ conversion and influence of these new processes for properties of obtained \ggam collider were absent in literature.  And these problems
are the main subject of the present paper.

We will show that in this approach  at electron beam energy up to 1 TeV one can
obtain reasonable high energy luminosity with the very narrow energy distribution. The laser system developed for PLC at the first stage of LC (for standard PLC) can be  applied here with the minimal variations (obliged mainly by change of the initial electron beam). To reach luminosity close to the best possible one, the
necessary laser flash energy should be enhanced by { about} 50\% only.

The organization of text is following.  Sections
\ref{sectechnintro} and \ref{secbasics} are introductory.
In the sect.~\ref{sectechnintro} we summarize main assumptions, technical
points and notations. Sect.~\ref{secbasics} is devoted to the
des\-crip\-tion of customary facts related to PLC at $x<4.8$ with
examples for the case $x=4.5$.

After that in the main body of
the paper we consider cases with $x=9$ and 18  which
correspond to  the initial electron beam energy
500 GeV and 1000~GeV (ILC, CLIC), and present some examples for $x=100$ (the electron beam energy 5~TeV) -- sect.~\ref{secbigx}.
We consider
the high energy part of spectra,  which is well separated from
their low energy part.  We start this description from the discussion
of process $\gamma\gamma_o\to\epe$ (``killing process'') in the conversion region (see notations below).
This process is absent at lower energies considered in earlier papers, i.e. at $x<4.8$. After a brief
discussion of the basic Compton effect
at $x>4.8$ (sect.~\ref{secbassp}), we
present equations for
balance of photons, produced in the Compton process and disappeared in the the conversion region  due to killing processes -- sect.~\ref{seceq}.

Next point is the optimal choice of the laser flash energy which correspond to the choice of the optical length  for the laser flash.  This choice is provided
the maximal number of photons for physical studies under
interest (sect.~\ref{secopt}).
 The sect.~\ref{secdistr} contains main results of the paper --  the description of high energy luminosity
spectra for \egam \ and \ggam \ collisions.

 Summary
(sect.~\ref{secsum})  presents  a brief description of the obtained
results. In this respect it is interesting to note  the following fact.
It is natural to expect that the shape of the resulting photon spectrum  will differ markedly from that of the basic Compton effect. To our surprise, the difference was not very large if we use
a reasonable choice of the polarization for the initial beams. It is due to fact that
the effects of the energy dependence are compensated by changing the polarization.

In Appendix~\ref{secbad} we discuss  case of ``bad'' choice of initial polarization.  In Appendix~\ref{seclinear}, we show that  the HE PLC studies with linear polarization of high energy photons
 are practically impossible.
Most important background is discussed in Appendix~\ref{appa}.
In Appendix~\ref{secphys} we briefly discusses New Physics problems that cannot be studied using LHC and \epe colliders and which can be studied using HE PLC.

\section{Introduction. Technical}\label{sectechnintro}

\paragraph{The principal assumptions and notations.}
For definiteness, we consider as a basic the $e^-e^-$ LC (not
 $e^+e^-$). We consider the effects of high photon density in the conversion region (nonlinear QED effects) to be negligible. \\
{\bf Notations.}
\begin{Itemize}
\item $\gamma_o$, $\omega_o$, $\lambda_o$ --   the  laser photon,  its energy and helicity;

\item $e_o$, $E $, $\lambda_e$ --  the  initial electron, its energy and helicity ($2|\lambda_e|\le 1$);

\item  $\gamma$, $\omega$, $\lambda$ --  the  produced photon, its energy and helicity;
\item $\sigma_0=\pi r_e^2=\;2.5\cdot 10^{-25}\;\mbox {cm} ^{2}$;

\item $b$ --  the  distance  between  the conversion  region $CR$ and  the
interaction region $IR$ (Fig.~\ref{fig:basschem});

\item   $\Lambda_C=2\lambda_e\lambda_o$ -- "polarization of   the process";
\item $y=\omega/E$
 --  relative photon energy;
\item $y_M=x/(x+1)$ --  maximal value of $y$ at given $x$;
\item
$w_{\gamma\gamma}=\fr{\sqrt{4\omega_1\omega_2}}{2E}\equiv \sqrt{y_1y_2}\le y_M$ and
 $w_{e\gamma}=\fr{\sqrt{4E\omega_1}}{2E}\equiv \sqrt{y_1}\le \sqrt{y_M}$ --
ratios
of the \ggam \  and \egam \ cms energies to $\sqrt{s}=2E$.
\vspace{5mm}
\end{Itemize}

We consider relative luminosities  $\cal L$, and  high energy luminosity integral
\beg
L_{exp}(w)={\cal L}(w)\cdot L_{geom},\\ {\cal L}_{h.e.}   =\!\!\int\limits_{y_{min}}^1 \!\!{\cal L}(w)dw\;\;\left(w=w_{e\gamma} \;\; or \;\; w_{\gamma\gamma}\right).\label{Lumintdef}
\eeg
The quantity $y_{min}$ here
is the  position of a minimum in the energy distribution of Compton photons at $x>3$  and $\Lambda_C\approx -1$, it depends on $x$ -- see \eqref{taudef} and Table~\ref{hepeak}. (The choice of  this quantity  does not require high accuracy, since the number of Compton photons at $y\approx y_{min}$ is small.)

Here $L_{geom}$ is  luminosity of the $e^-e^-$ collider
prepared for \ggam \ mode\fn{It is useful to take into account that the
standard beam collision effects absent for $\gamma \gamma$ collisions. Therefore, this $L_{geom}$  can be made
even larger than the anticipated luminosity of basic \epe
collider.}. In a nominal ILC option, i.e. at the  electron beam
energy of 250~GeV, the geometric luminosity  can reach  $L_{geom}=12\cdot 10^{34}$ cm$^{-2}$s$^{-1}$ which is
about 4 times greater then the anticipated \epe  luminosity.

All numerical calculations are carried out for the case of a "good" polarization of collided electrons and photons $\Lambda_C\approx -1$,
in two versions: $\Lambda_C= -1$ and $\Lambda_C= -0.86$  (more realistic case).

We distinguish luminosities $L_I$ with different total helicity\lb
$I=|\lambda_1-\lambda_2|$ of initial state,  that are ($L_{1/2}$ and $L_{3/2}$ for \egam collisions) and ($L_0$ and $L_2$ for \ggam \ collisions).
We find that one of these helicities dominates in the most important and useful high energy part of the luminosity spectrum.
Therefore, below we discuss
total luminosity (sum over both final helicities) and present
fraction of final state, giving smallest contribution into
luminosity.  Trivial change of sign of helicity of one electron and laser beam allows produce states with total helicity 2 instead 0 -- for \ggam \ collisions and 3/2 instead of 1/2  -- for \egam collisions.

\bu  The obtained luminosity spectra can be treated as the sum of
two separate distributions.

$\triangledown$ We are only interested in  {\bf the high-energy part} of the luminosity spectrum. It is formed by photons obtained in the basic Compton process \eqref{basComp}, some of these photons die in the processes \eqref{killing}.
An important feature of the Compton effect for  $x>3$ is as follows: {\it
with a suitable polarization of the colliding electrons and laser light, the high-energy part of the resulting photon spectrum is well separated from the low-energy part, and this separation is enhanced with increasing electron energy $E$.   The dependence of details of experimental device can be eliminated from description of this part \cite{GKrho} --  see discussion around Eq.~\eqref{rhodef}}.

$\triangledown$    {\bf The low-energy part} of photon spectra is formed by low energy photons from the basic Compton process \eqref{basComp} and photons   from the  scattering of laser photons from the tail of the laser bunch  {\it (i)} on  electron after the first Compton process    \eqref{rescat} or {\it (ii)} on positrons or electrons produced in the killing process \eqref{parasitic}.
The  shape of  the energy distribution  in this region  depends  on details of the experimental device.

\bu  Below, the laser bunch is assumed to be wide enough so that a change in its density inside the electron bunch can be neglected.

The optical length of   the laser bunch for electrons is expressed
via   the  longitudinal density of photons $n_L$ in flash { (that is via the laser bunch energy $A$ divided to its effective transverse cross section
 $S_L$)} and  the total
cross section of   the Compton scattering $\sigma_C(x,
\Lambda_C)$:
\be
z =\fr{A}{\omega_o S_L}\,\sigma_C(x,
\Lambda_C)\equiv\fr{A}{A_0}
\fr{\sigma_C(x, \Lambda_C)}{\sigma_C(x\!=\!4.5, \Lambda_C\!=\!-1)}\,.
\label{ZAcorresp}
\ee
In the last form of this equation we introduce $A_0$ -- the laser flash energy, necessary to
obtain $z=1$ at $x=4.5$, $\Lambda_C=-1$.

When  the electrons traverse
 the laser beam, their number decreases as
\be
n_e(z)=n_{eo}e^{-z}\,.\label{numbel}
\ee

\begin{table}[hbt]\begin{center}
\begin{tabular}{|c||c|c| c|c|}\hline
$x$&4.5&9&18&100\\\hline
$\Lambda_C\!=\!-1$&0.73&0.45&0.26&0.056\\\hline
$\Lambda_C=1$&0.85&0.63&0.44&0.145\\\hline
\end{tabular}
\caption{\it
Compton cross section $\sigma_C/\sigma_0$ at  different $x$ and polarizations}\end{center}
\label{tabCompxsec}
\end{table}

\bu For the most suitable  modern laser
with $\omega_o= 1.17$~eV  at\lb $E=250$~GeV we have  $x=4.5$. For the
next stages of LC projects with $E=500$~GeV and $1000$~GeV
we have $x=9$ and $x=18$ respectively. The process \eqref{killing},  killing  high energy photons, is switched on at
${x>2(1+\sqrt{2})\approx 4.8}$.

 \paragraph{Processes in the conversion region.}

Let us list {\bf  processes in the conversion region:}
\bes\label{processes}
\bea
e_o\gamma_o\to e\gamma\;&\mbox{\it basic Compton process};\label{basComp}\\
e\gamma_o\to e\gamma\;;&\mbox{\it rescattering}\label{rescat} \\
\gamma\gamma_o\to \epe;&\mbox{\it killing process};\label{killing}\\
e^\pm\gamma_o\to e^\pm \gamma\;;&\mbox{}\label{parasitic}\\
e_o\gamma_o\to e\epe\;&\mbox{\it Bethe-Heitler  process}\,.\label{BHproceq}
\eea\ees

\bu \  The  Bethe-Heitler  process \eqref{BHproceq},  switching on at $x=8$,  is the process of the next order in $\alpha$ but its cross section  does not decrease with growth of energy, in contrast to the Compton process:\\
\cl{$\sigma_{BH}=(28/9\pi)\alpha \sigma_0 (\ln x-109/42)\;\;$ at $x\gg 1$.}
\begin{table}[hbt]\begin{center}
\begin{tabular}{|c||c|c|c|}\hline
$x$ &30&100&300\\ \hline
$\Lambda_C=-1$ &0.96&0.80&0.47\\ \hline
$\Lambda_C=1$ &0.97&  0.88&0.62\\ \hline
\end{tabular}
\caption{\it Factor $K_{BH}$}\end{center} \label{tabBH}
\end{table}
Because of this process, the yield of high-energy photons decreases by the factor
\be
K_{BH}=\fr{\sigma_C(x)}{\sigma_C(x)+\sigma_{BH}(x)}\,,\label{BHfac}
\ee
and the \ggam \ luminosity reduces  by the factor $K_{BH}^2$.
The numerical values of this factor are presented  in the Table~\ref{tabBH}.
It shows that  the Bethe-Heitler mechanism is negligible at  $x<100$,  the reduction of the photon yield becomes unacceptably large at $x>300$.

\section{Basics}\label{secbasics}

Before main discussion, we repeat some basic points from papers
\cite{GKST}  in the form, suitable for description of high energy
part of spectra  at large $x$,  with addition of some details which were not discussed previously.

In this section  numerical examples are given for the standard
case $x=4.5$, which is close   to   the upper limit
of validity  in previous  studies. The photon energy is
kinematically bounded from above by quantity $y_M=x/(x+1)$. For $x=4.5$ we have $y\le y_M=0.82$.

The total cross section of Compton effect is well known (Table~\ref{tabCompxsec}):

\beg
\sigma_C=\fr{2\sigma_0}{x}\left[F(x)+\Lambda_C T(x)\right], \\
F(x)\!=\!\left(1-\fr{4}{ x}\! -\! \fr{8}{x^2}\right)\ln (x+1)\! +\!\fr{1}{2} \!+\!\fr{8}{x}\! -\! \fr{1}{ 2(x+1)^{2}} ,
 \\
T(x)\!=\!\left(1\!+\!\fr{2}{x}\right) \ln (x+1)\!-\!\fr{5}{2}\!+\!\fr{1}{ x+1}\!-\!\fr{1}{2(x+1)^{2}}.
  \label{xComp}\eeg

The photon energy distribution at $E\gg \omega_o$ is given by\\ ($r=\fr{y}{x(1-y)}\le 1$)
 \beg
f(x,y)\!\equiv \!\fr{1}{\sigma_
 C}\, \fr{d\sigma _C}{
dy}=\fr{U(x,y)}{F(x)+\Lambda_C T(x)}\equiv\\
\!\equiv\!\fr { \left[\fr{1}{1-y}\! +\! 1\!-\!y\! -\! 4r(1\!-\!r)\! -\!
\Lambda_C\, xr(2\!-\!y)\,(2r\!-\!1)\right]}{F(x)+\Lambda_C T(x)}.\label{yComp}
 \eeg

\begin{figure}[hbt]\begin{center}
\includegraphics[width=0.4\textwidth]{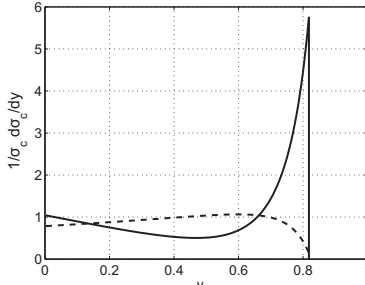}\vspace{-6mm}
\caption{\it Photon energy spectrum $f(x,y)$ at $x=4.5$, $\Lambda_C=-1$ (solid)
and $\Lambda_C=1$ (dotted)} \label{fig:Compsp48}\end{center}
\end{figure}
The shape of this distribution  depends strongly on both
$x$ and polarization of process. At $\Lambda_C\approx-1$,  the
photon spectrum grows up to its upper boundary $y=y_M$, at
$\Lambda_C\approx 1$ this spectrum is much more flat
(Fig.~\ref{fig:Compsp48}).

 At $x>3$ and $\Lambda_C\approx -1$ the high
energy part of this distribution is concentrated in the narrow strip below upper limit,  contained more than one half produced photons. We  characterize this peak by its lower boundary $y_{min}$ and sharpness parameter $\tau$ -- see Table \ref{hepeak}
\beg
\mbox{\it definition of $y_{min}$:}\;\; \quad \int\limits_{y_{min}}^{y_M}f(x,y)dy\approx 0.55;\\
f(x,y\le y_{min})\ll f(x,y_M)\,;\\
\mbox{\it definition of $\tau$:}\quad f(x,y=y_M(1-\tau))=\fr{1}{2}\cdot f(x,y_M)\,.\label{taudef}
\eeg
\begin{table}
\centering{
\begin{tabular}{|c||c|c|c|c|}\hline
x& 4.5&9&18&100\\\hline
$y_{min}$ &0.6&0.7&0.75&0.94\\
$\tau$&&0.036&0.022&\\\hline
\end{tabular}
\caption{\it Properties of high energy peak at $\Lambda_C\approx -1$}
\label{hepeak}}
\end{table}
The mean circular polarization (helicity) of   the produced photon is
\bes \label{polComp}\beg
 \lambda_C(y) =\\=\fr{
-\lambda_o(2r\!-\!1)[(1-y)^{-1}\!+\!1-\!y]\!+\!2\lambda_e
xr[1\!+\!(1\!-\!y)(2r\!-\!1)^2]}{U(x,y)}.\label{polComp1}
  \eeg
At $\lambda_o=\pm 1$, this equation is simplified:
\be
 \!\!\lambda_C(y)
\!=\!-\!\lambda_o\left[(2r\!-\!\!1)\!+\!\fr{
\left[(2r-1)-\Lambda_C xr\right]4r(1-r)} {U(x,y)}\right]\!.\label{polComp2}
  \ee
  \ees
The photons of highest energies  are well polarized with the same
direction of spin as laser photons, i. e. $ \lambda_C(y=y_M)
=-\lambda_o$.

The photons of maximal energy  move in the direction of the initial electron. With decreasing of
 the photon energy   its escape angle  grows as
 \be
\theta (y)=\fr{m }{E}\; \sqrt{x+1}\; \sqrt{\fr{y_M}{ y}-1}.
 \label{thety}
 \ee

The \ggam \ or \ \egam \ luminosities are given by convolution of
high energy photon spectrum with spectrum of collided photon or
electron taking into account   the mentioned spread of
photons at the way from  the conversion region $CR$ to the
interaction region $IR$. Owing to the  angular spread
\eqref{thety}, at this way  more soft photons spread for more wide
area, and collide more rare, as a result, their contribution to luminosity
decreases. Therefore, with the growth of   the distance $b$
between  the conversion region $CR$ and   the
interaction region $IR$ luminosity decreases but monochromaticity
improves.  {\it At $y>y_M/\sqrt{2}$} \ these
effects are described with good accuracy by   a
single parameter -- as for round beam
\cite{GKrho}
 \be
\rho^2=\left(\fr{b}{(E/m_e)\sigma_x}\right)^2+\left(\fr{b}{(E/m_e)\sigma_y}\right)^2\,.
\label{rhodef}
 \ee
(Here $\sigma_x$ and $\sigma_y$ are semiaxes of
ellipse, described initial electron beam in the {\it would be}
interaction point\fn{\it At lower photon energies
this one-parametric description is invalid due to both geometrical reasons and contribution of rescattering \eqref{rescat}.}.
In this approximation,  the discussed luminosities
are expressed via distributions in the photon energy \cite{GKST}:
\bea
&\fr{dL_{\gamma e}^{\lambda}}{dy}=n_e n_\lambda (y,z)
e^{-\fr{\rho^2\phi^2}{2}}\,\quad
\left(\phi_a=\sqrt{y_M/y_a-1}\right); &\\\label{Legam}
&\!\!\!\!\fr{dL_{\gamma\gamma}^{2,\lambda_1\lambda_2}}{dy_1dy_2}\!=\!
n_{\lambda_1} (y_1,z)n_{\lambda_2} (y_2,z) I_0
(\rho^2\phi_1\phi_2)e^{-\,\fr{\rho^2(\phi_1^2+\phi_2^2)}{2}}\,.&
 \label{Lggam}
 \eea
Here $I_0(z)$ is  the modified Bessel function. The distributions over the center of mass energy $w$ are obtained by the
substitution\lb $w=\sqrt{y}$ \ for \egam \ luminosity or
$w=\sqrt{y_1y_2}$ with simple integration, for \ggam \ luminosity.

At $x\le 4.8$ each missed electron produces the photon.
Therefore at $z=1$ and $\rho=0$ \  total \egam \ and \ggam \ luminosities are
\be
{\cal L}_{e\gamma} =(1-1/e)=0.63\,,\qquad
{\cal L}_{\gamma\gamma} =(1-1/e)^2=0.4.\label{lumz1}
\ee

\begin{figure}[h!]
\centering{\includegraphics[width=0.4\textwidth,height=0.25\textheight]{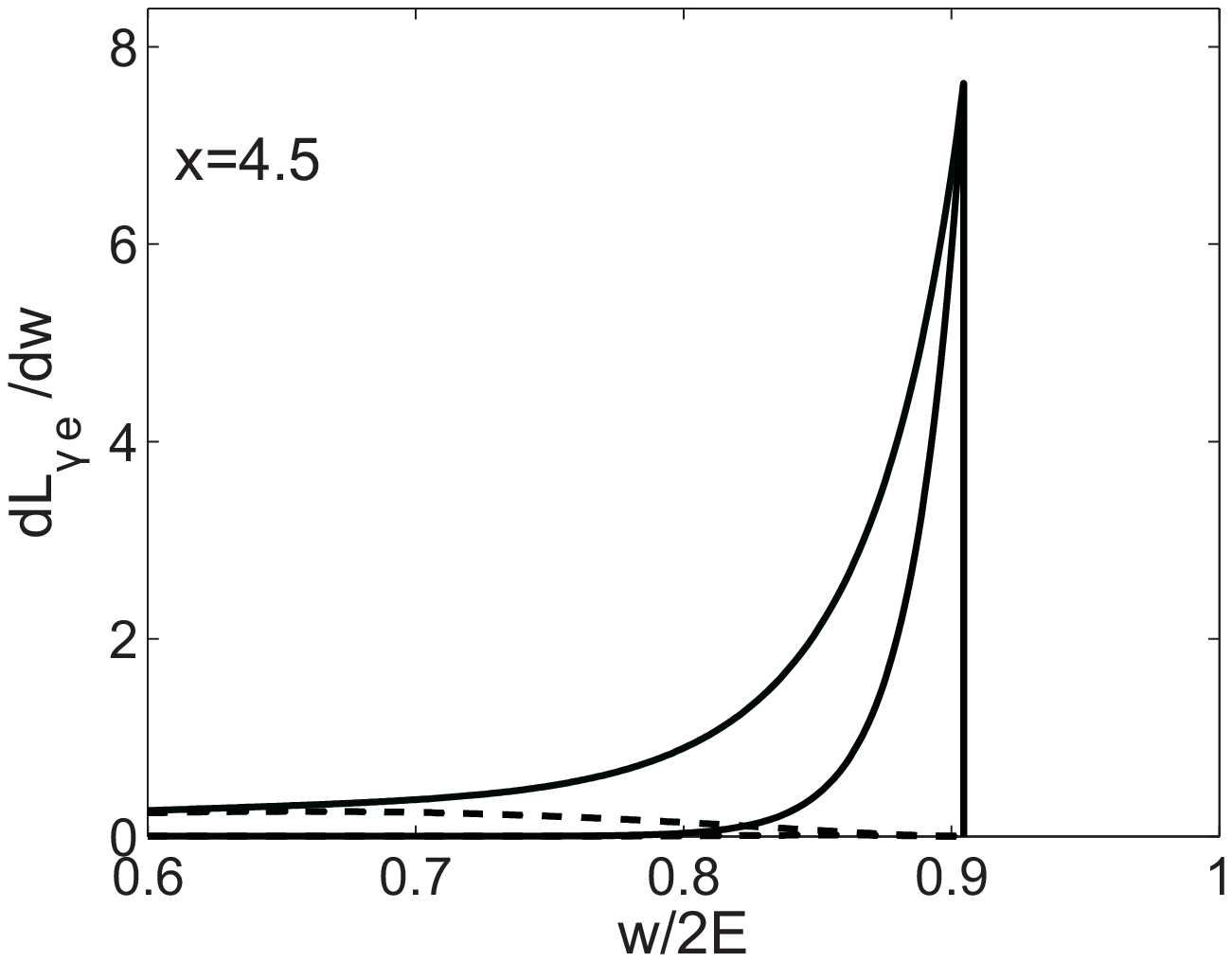}\hspace{5mm}
\includegraphics[width=0.4\textwidth,height=0.25\textheight]{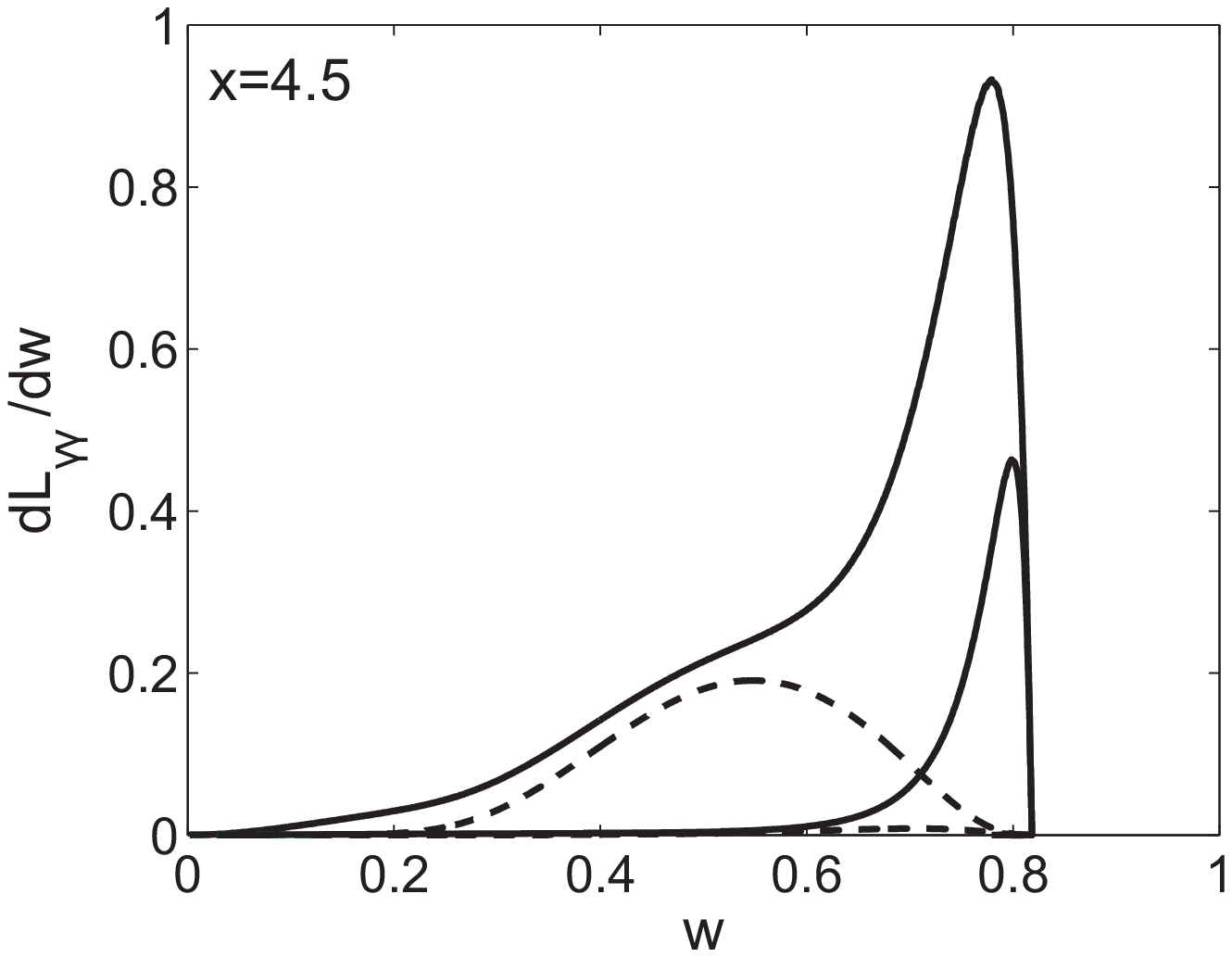}}  
\caption{\it Luminosity spectra $dL/dw$ at $x=4.5$, $\Lambda_C=-1$, $z=1$  for $\rho=1$ (up) and $\rho=5$ (down) -- solid lines, dotted lines -- similar distributions at $\rho=1$ for final states with total helicity $3/2$  for \egam \ collisions (left) and total helicity 2 for \ggam \ collisions ( right).}
  \label{fig:lumsp45}
\end{figure}

 Fig.~\ref{fig:lumsp45} and first lines of
Tables~\ref{tabggamlumLam1}, \ref{tabegamlum45918} represent \ggam
\   and \ \egam \ luminosity distributions in their cms  energy
at  $\Lambda_C=-1$,  $z=1$ for $\rho=1$ and 5.

At $\Lambda_C=-1$  and $\rho\ge 1$,  the  main part of
calculated luminosity is concentrated at $w>0.6$ (in particular,
the \egam \ luminosity spectrum at $y>0.6$ contains about 2/3 of
produced photons). The relative values of luminosity with total
helicity 2 for \ggam \ collisions and luminosity with total
helicity 3/2 for \egam \ collisions are small and decrease
quickly with growth of $\rho$, it means that the colliding
photons become practically monochromatic in their polarization.

Therefore,  we  describe the luminosity distribution by
three parameters:\\ 1)  the
integral ${\cal{L}}_{h.e.}$ \eqref{Lumintdef},\\  2) the
relative value of contribution with total helicity 2 or $3/2$,\\
3)  position of maximum
$w_M$ in summary luminosity  and corresponding maximal value
$L(w_M)$. (For \egam \ collisions peak is
disposed at  the maximal value of $w=\sqrt{y_M}$, and
$L(w_M)$ is independent on $\rho$).

\section{At $\pmb{x\ge 4.8}$}\label{secbigx}

In principle, all processes \eqref{processes} should be taken into account.
Only two of them, basic Compton effect \eqref{basComp} (discussed in previous section) and killing process \eqref{killing} take part in formation of high energy part of photon spectrum under interest.
The  process \eqref{parasitic} adds low energy photons like  the
process \eqref{rescat}. The contribution of these processes  into high energy part of photon spectrum is  small       at considered  moderate conversion coefficients.

\subsection{Killing process \eqref{killing}}

The killing process describes the disappearance of a Compton high-energy photon in its collision with a laser photon from the tail of a bunch, generating
\epe \ pair, it switches on at  $x\approx 4.8$.  For the photon with energy $yE$ the squared cms  energy for process is $w^2_k=4\omega\omega_o/m_e^2=xy>4$. Its cross section
is
\beg
\sigma_{kill}(w^2_k,\lambda_o\lambda)=\fr{4\sigma_0}{w^2}\Phi_{\ggam}(w^2_k,\lambda_o\lambda)\,,\\ \Phi_{\ggam}(w^2_k,\lambda_o\lambda)=
\left(1+\fr{4}{w^2_k}-\fr{8}{w^4_k}\right)L-\\
-\left(1+\fr{4}{w^2_k}\right)v-
\lambda_o\lambda(L\!-\!3v)
,\\
v=\sqrt{1-4/w^2_k\;},\quad L=2\ln\left(\fr{w_k}{2}(1+v)\right);\\ w^2_k=\fr{4\omega_o\omega }{m_e^2}\equiv xy\,.
\label{killproc}\eeg
Note that\fn
{\it At $\Lambda_C\approx -1$   in the high energy part of
spectrum we have $\lambda\approx -\lambda_o$. At $x>20$, $\Lambda_C\approx -1$  the last term  in $\Phi_{\ggam}$ provides faster  decreasing the number of high energy photons than  in the case of "bad" polarization $\Lambda_C\approx 1$.}
\beg
\fr{\sigma_{kill}(\lambda_o\lambda>0)}{\sigma_{kill}(\lambda_o\lambda<0)}>1 \;\; \mbox{at}\;\;w^2_k<15;\\
\fr{\sigma_{kill}(\lambda_o\lambda>0)}{\sigma_{kill}(\lambda_o\lambda<0)}<1 \;\; \mbox{at}\;\;w^2_k>15.
\eeg

\subsection{Basic spectra}\label{secbassp}

\begin{figure}[hb]
\centering{\includegraphics[width=0.4\textwidth]{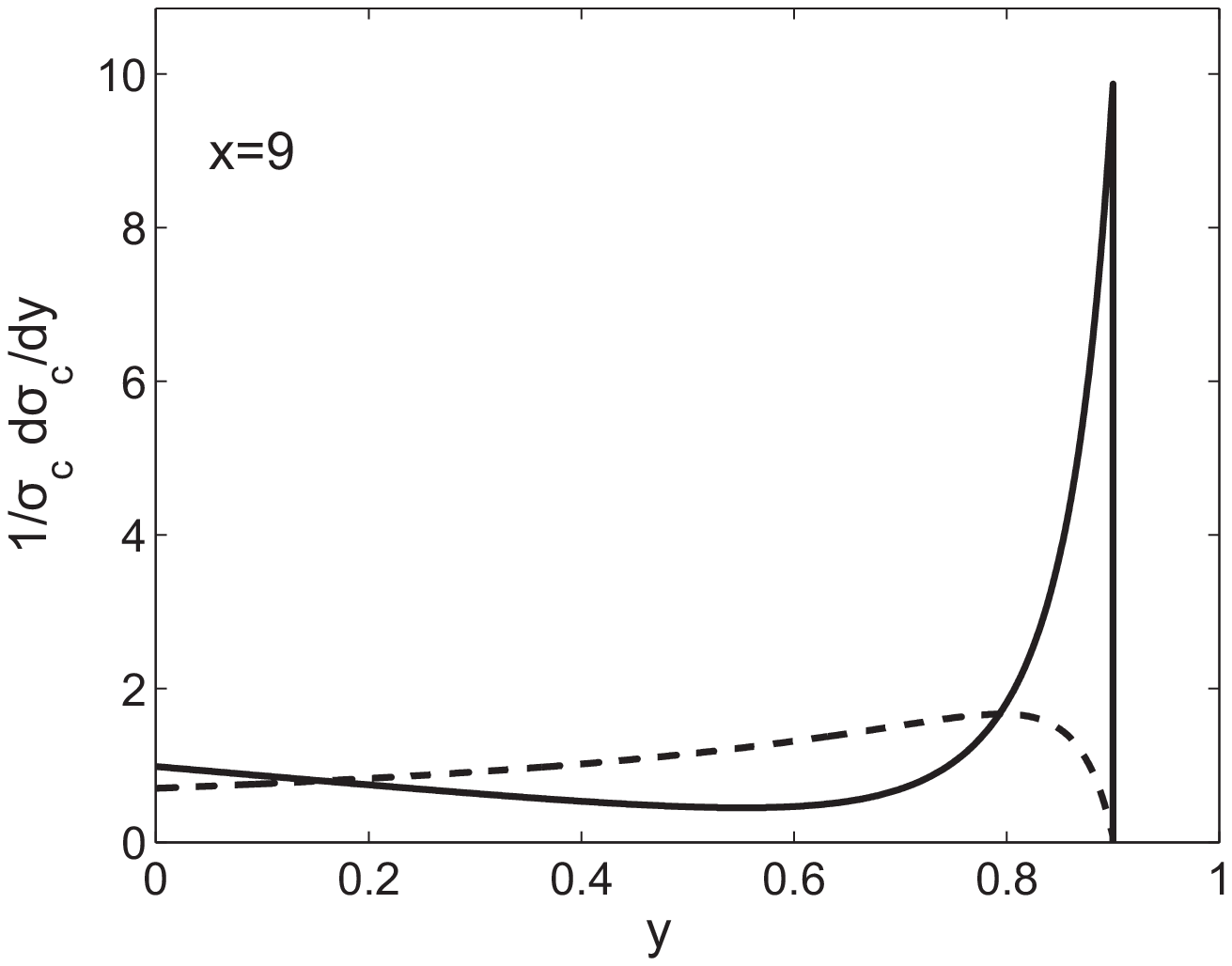}
\includegraphics[width=0.4\textwidth]{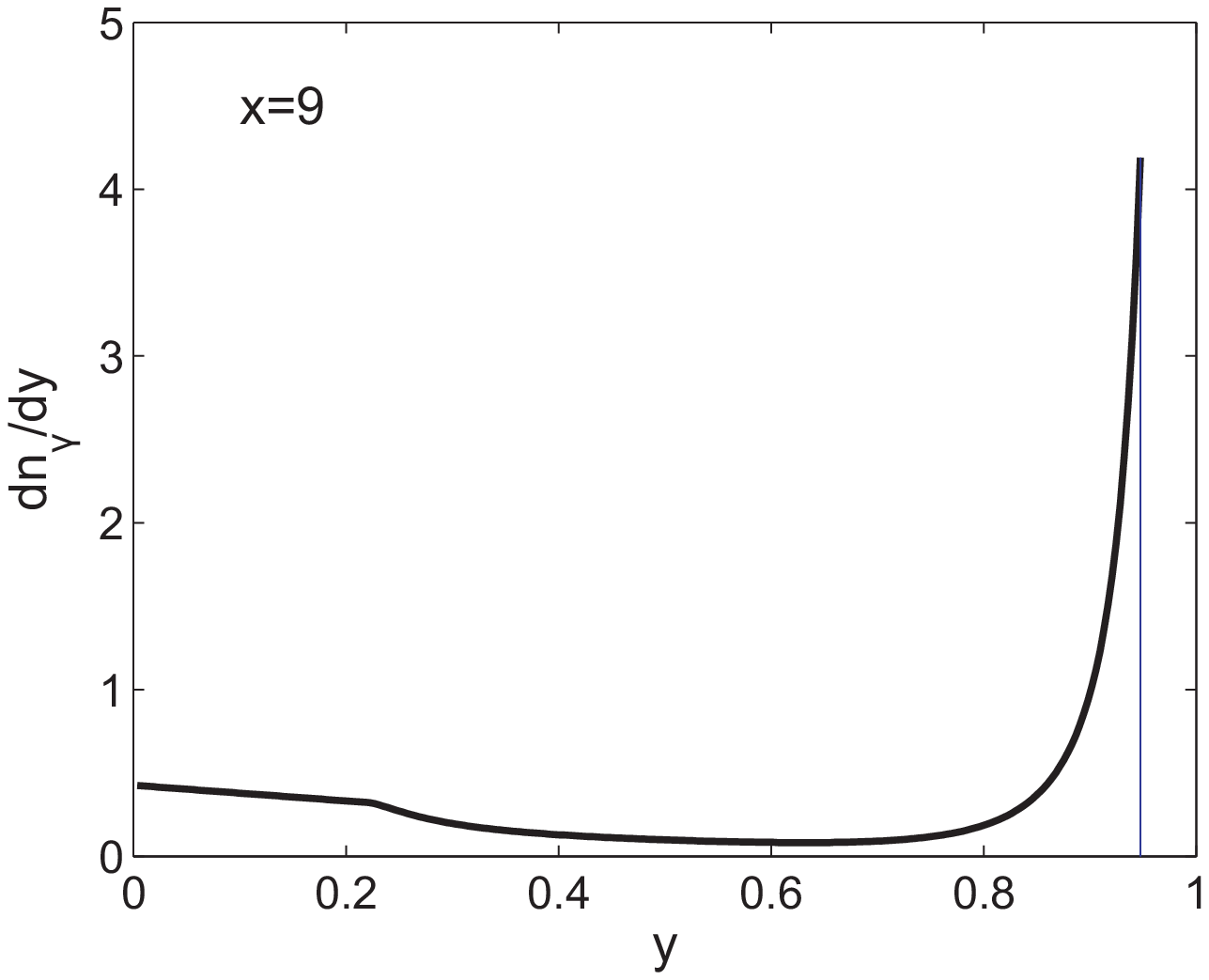}}
\caption{\it The case $x=9$. 
 Left -- Compton spectrum of photons.
Solid line corresponds  to  $\Lambda_C=-1$,
dashed line --  to $\Lambda_C=1$. 
Right --
photon energy spectrum  reduced by
killing process at $\Lambda_C=-1$, $z=0.7$} \label{x918spectra}
\end{figure}

The photon energy spectra for the basic Compton
backscattering \eqref{yComp}  at $x=9$  is shown  in
Fig.~\ref{x918spectra} -- left.

It is clearly seen that this spectrum for $\Lambda_C=-1$ is
concentrated near the high energy limit  strongly then the
corresponding spectrum for  $x=4.5$ in
Fig.~\ref{fig:Compsp48}.

At $x=18$ similar calculations show that the
spectrum is concentrated in the  narrow strip $0.85<y\equiv
\omega/E <0.95$. At $\Lambda_C=1$ these spectra are much more
uniform, without  the big  peak  at high energies.

\subsection{Equations}\label{seceq}

The balance of  high energy photons is given by their
production in the Compton process \eqref{basComp} and their
disappearance at the production of \epe \ pairs   in the
collision of these photons with residual laser photon (killing process \eqref{killing}).

Let us denote by $n_\gamma(y,z,\lambda)$ the flux of photons (per
1 electron) with  the energy $yE$ and polarization $\lambda$
after travelling  inside the laser beam with  the
optical length $z$. It is useful for calculations to decompose this photon flux to
the sum of fluxes of right polarized photons $n_{(\gamma +)}(y,z)$
and left polarized photons $n_{(\gamma -)}(y,z)$
so  that the total photon flux $n_\gamma(y,z)$ and its average
polarization $\lambda$ are
 \beg
 n_\gamma(y,z)=n_{(\gamma +)}(y,z)+n_{(\gamma -)}(y,z), \\ \la\lambda(y,z)\ra=\fr{n_{(\gamma +)}(y,z)-n_{(\gamma -)}(y,z)}{n_{(\gamma +)}(y,z)+n_{(\gamma -)}(y,z)}\,.\label{densdef}
\eeg
Naturally,  $\la\lambda(y, z\to 0)\ra\to\Lambda_C(y)$ from \eqref{polComp}.

The variation of these fluxes  during the travelling
inside the laser bunch is described by equations
\beg
\fr{dn_{(\gamma\pm)}(y,z)}{dz}=\\\!=\!\fr{1}{2}\left(1\pm
\Lambda_C(y)\right) f(x,y)n_e(z)\! -\!
n_{(\gamma\pm)}\fr{\sigma_{kill}(xy,
\pm\lambda_o)}{\sigma_C(x)}.\label{eqdens}
 \eeg
The number of photons and their mean polarization are expressed
via subsidiary quantities $\nu_\pm$:
 \ba{c}
n_\gamma(y,z)=f(x,y)\nu(z,y),\\ \nu(z,y)=
\nu_+(z,y)+\nu_-(z,y),\quad
\la\lambda\ra=\fr{\nu_+(z,y)\!-\!\nu_-(z,y)}{\nu(z,y)}.
\ea
The equation \eqref{eqdens} is easily solved:
 \beg
n_{(\gamma\pm)}(y,z)=f(x,y)n_{e0}\nu_\pm(y,z);\\ \mbox{where}\;\;
\nu_\pm(y,z)=
\fr{\left(1\pm \Lambda_C(y)\right)}{2}
\cdot\fr{e^{-\zeta_\pm z}- e^{-z}}{1-\zeta_\pm};\\
\zeta_\pm= \fr{\sigma_{kill}(xy, \pm\lambda_o)}{\sigma_C(x)} \,.
\label{solveq}\eeg

It is useful for future discussions to define in addition ratio of number of killed photons to the number of  photons, prepared for the \ggam \ collisions, $r_K(y,z)$ -- see Table~\ref{1bbb}.

In Fig.~~\ref{x918spectra} we compare photon energy spectrum for
the pure Compton effect (left plot) and after travelling
inside the laser bunch with $z=0.7$ (right plot). One
can see  at the right plot that (1)  the shape of the high-energy part of the spectrum reproduces approximately that for the pure Compton effect; (2)
in the middle part of
spectrum the killing process leads to relative decrease of
spectrum; (3) the part of spectrum, correspondent to $xy<4$
is relatively enhanced, since at these $xy$
the killing process is absent.

\section{Optimization}\label{secopt}

At $x<4.8$,  the growth of  the optical length results in monotonous increase
of number of photons (with  the simultaneous growth of background).
On the contrary, at $x>4.8$, the killing process stops this increase and
even kills all high energy photons at extremely large $z$. The
dependence $n(y,z)$ on $z$  has maximum at some $z=z_m(y)$, which
can be treated as  the optimal value of $z$. However, what value of $y$ should we use?

The simplest approach is to consider this balance only for high energy photons, at $y=y_M$ \cite{Tel2001}.
We find it more reasonable to consider for this goal  the $z$-dependence of
entire luminosity  within its high energy peak ${\cal L}_{h.e.}$ \eqref{Lumintdef}, \eqref{taudef}, Table~\ref{hepeak}.

\begin{figure}[hbt]\centering{
\includegraphics[width=0.45\textwidth]{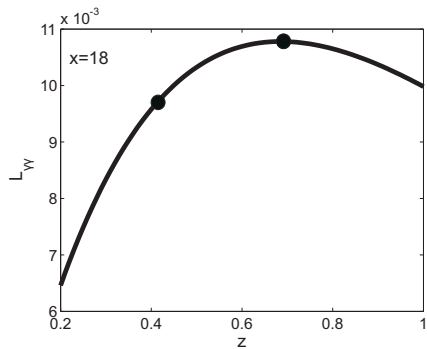}
\caption{\it The \ggam \ luminosity integral ${\cal L}_{h.e.}$ in dependence on $z$,
for $\rho=1$, $\Lambda_C=-1$. Dots correspond $z_m$ and $z_{0.9}$.} \label{figlumz}}
\end{figure}
The  typical dependence  of  luminosity ${\cal L}_{h.e.}$ on $z$  is shown in   Fig.~\ref{figlumz}.
The curves at another $x$ and $\rho$ have similar form\fn{\it At $\Lambda_C\approx -1$,  the photon energy spectra are concentrated
in the narrow region near the upper boundary $y=y_M$. Therefore, the results of both optimizations
are close to each other (in our examples  the difference in $z_m$ for two methods is less than 2\%).
At $\Lambda_C=1$ the initial spectra are much more flat in $y$, and value $z_m$ depends
on $\rho$ stronger, in this case estimates done for $y=y_M$ \cite{Tel2001}  cannot be used
for description of luminosity.}.
The
optimal value  of the laser optical length are given by the position of a maximum at these curves, $z_m$
They are shown in the table~\ref{1bbb}. We find numerically that the dependence of $z_m$ on $\rho$ is negligibly weak.

The curves like Fig.~\ref{figlumz} are very flat near maximum.
Therefore, value $z_{0.9}(x)<z_m(x)$, provided luminosity which
is 10\% lower than maximal one,  is noticeably less than $z_m$.
In the table~\ref{1bbb} we present for $\Lambda_C=-1$\\ {\it (i)} values $z_m$ and $z_{0.9}$;\\
{\it (ii)} laser flush energies \eqref{ZAcorresp} required
to obtain either maximal luminosity $A(z_m)$ or 90\% \ of this
luminosity $A(z_{0.9})$ (in terms of  the laser flush energy $A_0$, required
to obtain $z=1$ at $x=4.5$);\\ {\it (iii)} the proportion of killed photons among all photons born in the basic Compton effect (at $y=y_M$) $r_K(z,y)$.

\begin{table}[ht]
\begin{center}
 \begin{tabular}{||c|c||c|c|c|c|c|c|c|}\hline
x&$y_{min}$&z&$r_K(z,y_M)$&$A(z)/A_0$\\\hline
9&0.7&$z_m=0.704$&0.22&1.15\\ \cline{3-5}
&&$z_{0.9}=0.49$&0.13&0.8\\\hline
18&0.75&$z_m=0.609$&0.43&1.7 \\ \cline{3-5}
&&$z_{0.9}=0.418$&0.28&1.17\\\hline
100&0.94&$z_m=0.48$&&6.3\\\cline{3-5}
&&$z_{0.9}=0.32$&&4.2\\\hline
\end{tabular}
\caption{\it Optical lengths $z_m$ and $z_{0.9}$, fraction of killed photons and
necessary laser flash energy at these $z$ for $\Lambda_C=-1$}
\label{1bbb}
\end{center}
\end{table}

\section{Luminosity distributions}\label{secdistr}

\begin{figure}[hbt]
\includegraphics[width=0.48\textwidth,height=0.25\textheight]{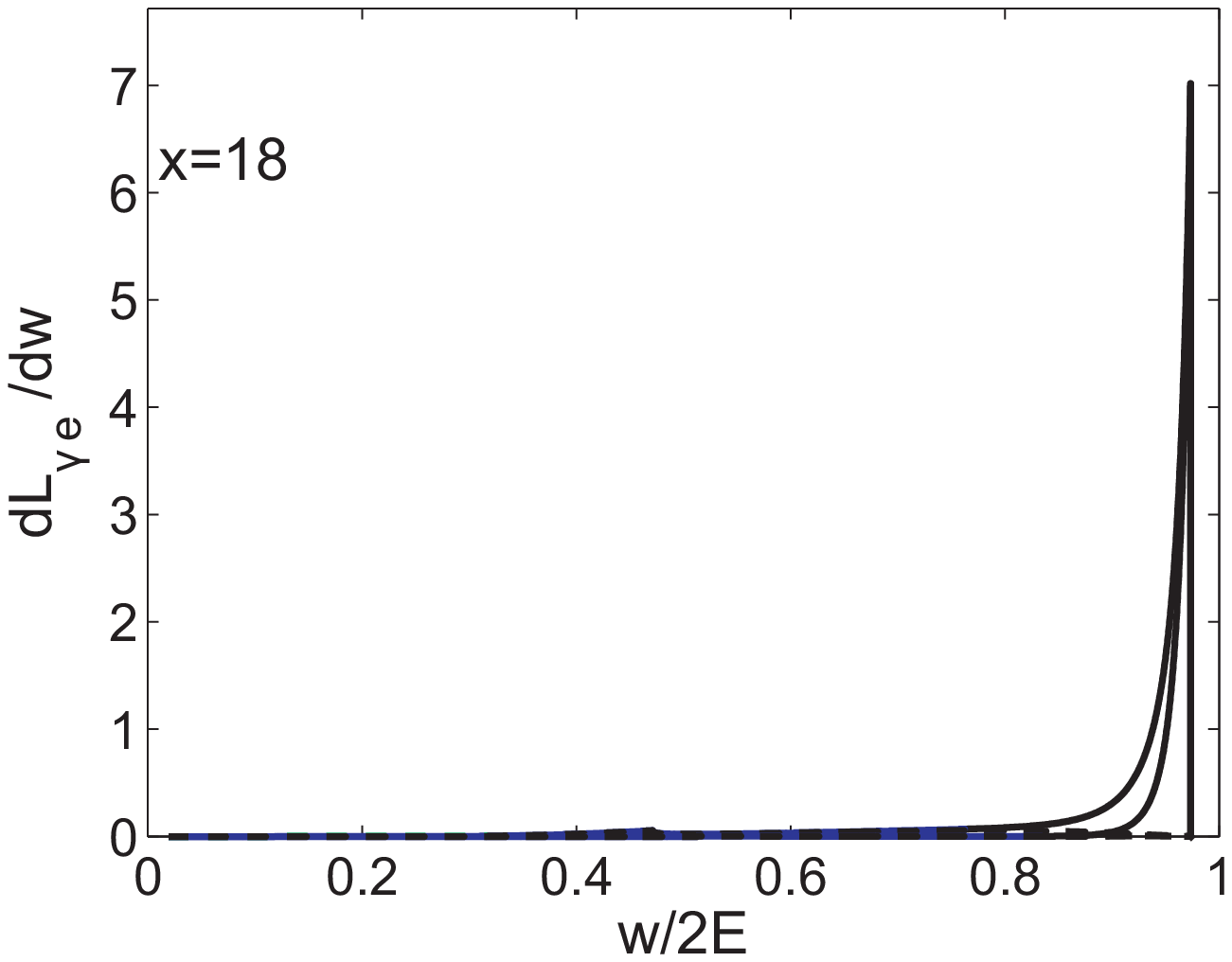}\hspace{5mm}
\includegraphics[width=0.48\textwidth,height=0.25\textheight]{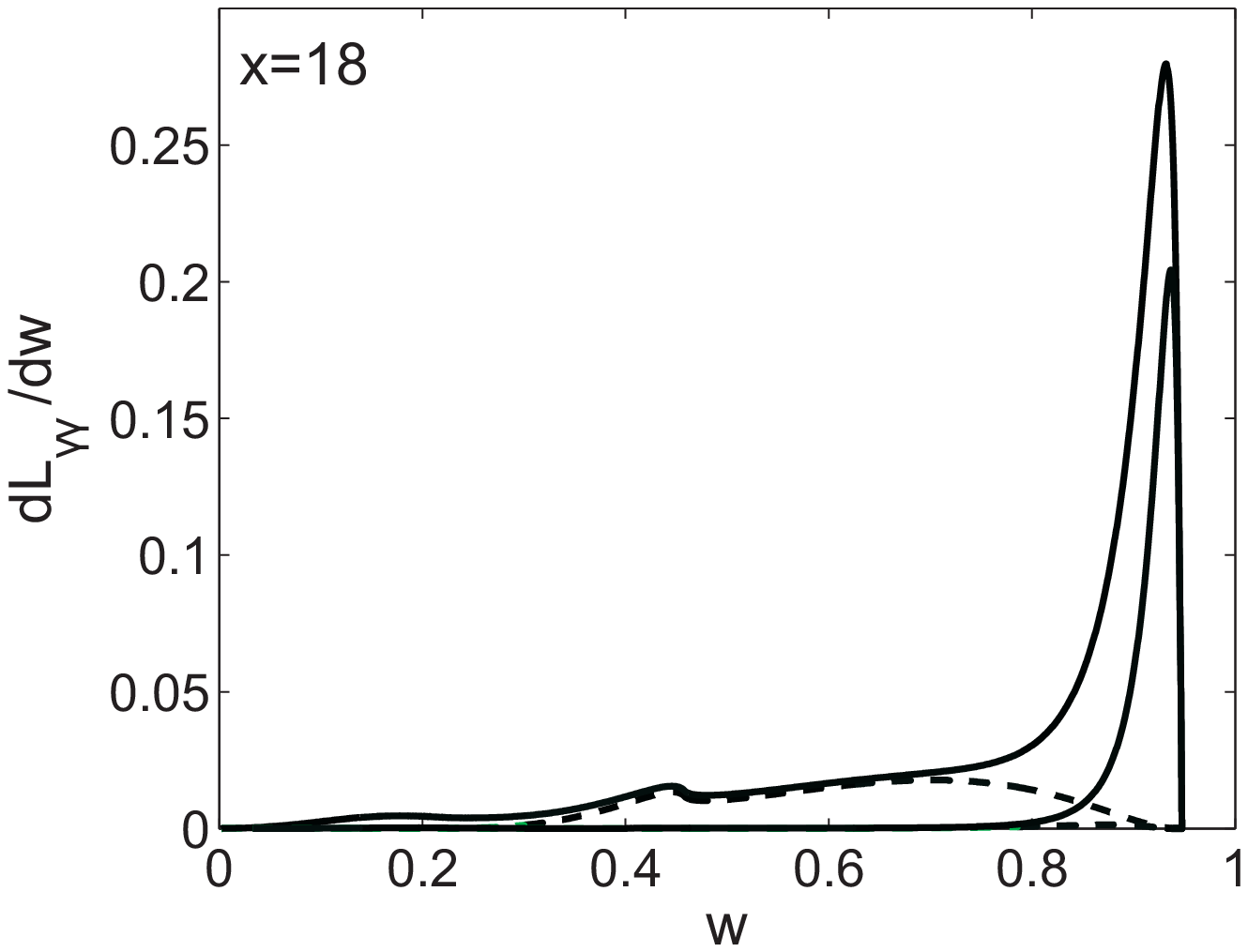}  
\caption{\it Luminosity spectra $dL/dw$ at $x=18$, $\Lambda_C=-1$, $z=1$  for $\rho=1$ (up) and $\rho=5$ (down) -- solid lines; dotted lines -- similar distributions for final states with total helicity $3/2$   ($\gamma e$ \ collisions) or 2 (\ggam \ collisions).}
  \label{lum18}
\end{figure}

Now we consider luminosity spectra for optimal $z=z_m$.
The examples of these spectra for \ggam \ and \egam \ luminosities at $x=18$, $\Lambda_C=-1$ for $\rho=1$ and 5 are shown in Fig.~\ref{lum18}. The curves for other $x$ and $\rho$ have similar form.

\bu The table~\ref{tabggamlumLam1}  represent main properties of high energy
peaks in the \ggam \  luminosity  spectra at $x=9,\,18$ for $\Lambda_C=-1$ and $\Lambda_C=-0.86$ at optimal $z=z_m$.
 Lines for $x=4.5$, $z=1$ and $x=100$, $z=z_m$ are presented  for comparison.
 Here \\
${\cal L}_{h.e.}$ --- total luminosity of the high energy peak \eqref{Lumintdef}, Table~\ref{hepeak},\\
$L_m= L(w_M)$ -- maximal differential luminosity,\\ $w_M$ -- position of  maximum in luminosity,\\
 $w_\pm$ -- solutions of equation  $L(w_\pm)=L(w_M)/2$,\\
$\gamma_w=(w_+-w_-)/w_M$ --  relative width of obtained peak.\\
$L_2/L$ -- fraction of   non-leading total helicity 2.


 \begin{table}[ht]
\begin{center} \begin{tabular}{|c|c|c|c|c|c|c|c |c|c|c|c|}\hline
$\rho$& $\Lambda_C$&${\cal L}_{h.e.}$&$L_2/L$&$L_m$&$w_M$&$w_-$&$w_+$&$\gamma_w$ \\\hline
\multicolumn{9}{|c|}{$\ggam: \qquad x=4.5,\qquad z=1,\;\;$}\\\hline
$\ba{c} 1\\ \\5\ea$&$\ba{c}-1\\-0.86\\\hline-1\\-0.86\ea$&
$\ba{c} 0.121\\0.114\\0.031\\0.0275\ea$&$\ba{c} 0.143\\0.215\\0.041\\0.086\ea$&$\ba{c} 0.933\\0.82\\0.464\\0.40\ea$
&$\ba{c}0.779\\0.778\\0.799\\0.7985\ea$&$\ba{c}0.689\\0.672\\0.760\\0.0758\ea$&$\ba{c}0.89
\\0.809\\0.814\\0.813\ea$&$\ba{c}0.154\\0.177\\0.067\\0.069
\ea$\\\hline\hline\hline

\multicolumn{9}{|c|}{$\ggam:\qquad x=9,\qquad z=z_m=0.704,\;\;$}\\\hline

$\ba{c} 1\\ \\5\ea$&$\ba{c}-1\\-0.86\\\hline-1\\-0.86\ea$&$\ba{c}0.0214\\0.0201\\0.0072\\0.064
\ea$&$\ba{c} 0.079\\0.164\\0.021\\0.074\ea$&$\ba{c} 0.222\\0.195\\0.137\\0.118\ea$
&$\ba{c} 0.872\\0.871\\0.885\\0.885\ea$&$\ba{c} 0.814\\0.806\\0.854\\0.852\ea$&$\ba{c} 0.894\\0.894\\0.896\\0.896\ea$&$\ba{c} 0.092\\0.100\\0.048\\0.050\ea$\\\hline\hline\hline

\multicolumn{9}{|c|}{$\ggam: \qquad x=18,\qquad z=z_m=0.609,\;\;$}\\\hline
$\ba{c} 1\\5\ea$&$\ba{c}-1\\-0.86\\\hline-1\\-0.86\ea$&$\ba{c}0.0178\\0.0178\\0.0074\\0.069
\ea$&$\ba{c} 0.089\\0.228\\0.021\\0.144\ea$
&$\ba{c} 0.2615\\0.229\\0.190\\0.164\ea$&$\ba{c} 0.9317\\0.931\\0.9365\\0.936\ea$&$\ba{c} 0.8932\\0.886\\0.9138\\0.912\ea$&$\ba{c} 0.9436\\0.09436\\0.9447\\0.09447\ea$&$\ba{c} 0.054\\0.062\\0.033\\0.035\ea$\\\hline\hline
\hline

\multicolumn{9}{|c|}{$\ggam: \qquad x=100,\qquad z=z_m=0.477,\;\;$}\\\hline
$\ba{c} 1 \\5\ea$&-1&$\ba{c}0.0093\\0.0070
\ea$&$\ba{c} 0.017\\0.009\ea$
&$\ba{c} 0.527\\0.519\ea$
&$\ba{c} 0.9867\\0.9867\ea$&$\ba{c} 0.9771\\0.9793\ea$&$\ba{c} 0.9890\\0.9890\ea$&$\ba{c} 0.012\\0.0097\ea$\\\hline
\hline

\end{tabular}
\caption{\it
Properties of high energy \ggam \ luminosity for different $x$ and $\Lambda_C$.  }
\label{tabggamlumLam1}
\end{center}\end{table}

{\it The specific features of these  distributions at
$x=9, \, 18$ are following:}
\begin{Enumerate}
\item Luminosity within peak at $\rho=1$ is about 2\% from $L_{geom}$, about ($5.5\div 7$) times less than that for $x=4.5$.
\item Peaked differential luminosity $L_m$  at $\rho=1$ is about 1/4 from that for $x=4.5$.
\item Both integrated  luminosity and its peak value decrease with the growth of $\rho$, but
slower  than at $x=4.5$.
\end{Enumerate}

\bu The table \ref{tabegamlum45918}  shows similar properties for the high energy peak in the \egam \ luminosity
spectra for ideal case $\Lambda_C=-1$ with three  changes:   $L_2\to L_{3/2}$,  $w_+\equiv w_M=\sqrt{y_M}$,  $\gamma_w=1-w_-/w_M$.
\begin{table}[ht]\begin{center}
 \begin{tabular}{|c||c|c|c|c||c|c |c|c|}\hline
$\rho$&${\cal L}_{h.e.}$&$L_m$&$L_{3/2}/L$&$\gamma_w$
&${\cal L}(W)$&$L_m$&$L_{3/2}/L$&$\gamma_w$\\\hline
\multicolumn{9}{|c|}{$x=4.5,\qquad z=1, \qquad w_M=0.9045$}
\\\hline\hline

$\ba{c} 1\\5\ea$&$\ba{c} 0.38\\0.143\ea$&7.626&
$\ba{c} 0.135\\0.007\ea$&$\ba{c} 0.0296\\0.0141\ea$&
$\ba{c} 0.372\\0.134\ea$&7.034&$\ba{c} 0.182\\0.079\ea$&$\ba{c} 0.031\\0.014\ea$\\\hline
&
\multicolumn{4}{|c||}{$\Lambda_C=-1$}&\multicolumn{4}{c|}{$\Lambda_C=-0.86$}\\ \hline\hline\hline

\multicolumn{9}{|c|}{$x=9,\qquad z=0.7047, \qquad w_M=0.949$}
\\\hline\hline
$\ba{c} 1\\5\ea$&$\ba{c} 0.153\\0.074\ea$&4.837&$\ba{c} 0.069\\0.05\ea$&$\ba{c} 0.0176\\0.0105\ea$&
$\ba{c} 0.149\\0.068\ea$&4.32&$\ba{c} 0.116\\ 0.063\ea$&$\ba{c} 0.018\\0.011\ea$\\\hline&
\multicolumn{4}{|c||}{$\Lambda_C=-1$}&\multicolumn{4}{c|}{$\Lambda_C=-0.86$} \\\hline\hline\hline

\multicolumn{9}{|c|}{$x=18,\qquad z=0.609, \qquad w_M=0.973$}
\\\hline\hline

$\ba{c} 1\\5\ea$&$\ba{c} 0.136\\0.08\ea$&7.015&$\ba{c} 0.0625\\0.0059\ea$&$\ba{c} 0.0098\\0.0070\ea$&
$\ba{c} 0.138\\0.076\ea$&6.439&$\ba{c} 0.12\\0.077\ea$&$\ba{c} 0.01\\0.007\ea$\\\hline
&
\multicolumn{4}{|c||}{$\Lambda_C=-1$}&\multicolumn{4}{c|}{$\Lambda_C=-0.86$}\\ \hline\hline\hline

\multicolumn{9}{c|}{$x=100,\qquad z=0.477, \qquad w_M=0.995$}
\\\hline\hline

$\ba{c} 1\\5\ea$&$\ba{c} 0.099\\0.083\ea$&22.905&
$\ba{c} 0.0086\\0.0042\ea$&$\ba{c} 0.002\\0.0018\ea$&
$\ba{c} 0.1085\\0.088\ea$&21.905&$\ba{c} 0.108\\0.103\ea$&$\ba{c} 0.002\\0.0018\ea$\\ \hline
&
\multicolumn{4}{|c||}{$\Lambda_C=-1$}&\multicolumn{4}{c|}{$\Lambda_C=-0.86$}\\ \hline\hline
\end{tabular}
\caption{\it
Properties of high energy \ \  \egam \ luminosity for different $x$ and $\Lambda_C$.  }
\label{tabegamlum45918}
\end{center}
\end{table}

{\it The specific features of these  distributions at
$x=9, \, 18$ are following}
\begin{Enumerate}

\item Luminosity within peak at $\rho=1$ is  about 15\%
from $L_{geom}$, about 2.5 times less than that for $x=4.5$ (for \egam). It decreases with the growth
of $\rho$.
\item Peaked differential luminosity $L_m$  is independent on $\rho$, it is
of the same order of value as  that for $x=4.5$.
\end{Enumerate}\vspace{5mm}

\bu {\it The features  which are common for \ggam \ and \egam \ mode:}
\begin{Enumerate}

\item  The  rapidity of produced \egam and \ggam systems relative to the rest frame of collider is contained within a  narrow interval, determined by the spread of photon energies within high-energy peak \eqref{taudef},
    \be
    \eta_{e\gamma}\in\left(\fr{1}{2(x+1)},\; \fr{1}{2(x+1)}+\bar{\tau}\right),\;\;
    |\eta_{\gamma\gamma}|\le\bar{\tau},\;\;with \;\; \bar{\tau}\approx\fr{\tau}{2}\,.\label{rapidggam}
    \ee

\item Photons within considered peak are well polarized. The ratio of smaller luminosity
${\cal L}_2$ or  ${\cal L}_{3/2}$  to the total one is small.
\item  At higher $\rho$  the collisions are more  monochromatic in both energy and polarization (at $\rho=5$  the contribution $L_2$ (or  ${\cal L}_{3/2}$) disappears practically).
\item The energy distributions of luminosity are very narrow, for \ggam \ collisions the peak
widthes $\gamma_w$ are comparable with those  for basic \epe \ mode, taking into account initial state radiation (ISR) and beamstrahlung (BS). For \egam \ collisions these distributions are more monochromatic than basic \epe \
collisions, taking in account ISR and BS.
\item The non-ideal polarization of  the  initial electron $2\lambda_e=-0.86$ instead of -1 reduces
 quality of spectrum not very strong.
\end{Enumerate}

\section{Summary }\label{secsum}

1. LC with  the electron beam energy $E\le 1000$~TeV allows to construct the high energy photon collider (HE PLC)
using the same lasers and optical systems as those designed for
construction PLC at $E\le 250$~GeV.

2. In contrast with  the  well described case $x<4.8$, the  increase of the laser flash energy
above the optimal value results not in increase, but decrease of high energy \ggam \ or \egam \ luminosity.

3. In HE PLC  with the electron beam energy $E=1$~TeV, the maximal photon energy is $\omega_m=0.95$~TeV ($\sqrt{s_{\ggam} }\approx 1.9$~TeV),
the \ggam  \ luminosity distribution is concentrated near  the upper  boundary with a spread of
5\%, in addition, almost all photons have the same helicity (+1 or -1 -- in accordance with the experimentalist choice).
The total luminosity integral for this high energy part of spectrum is  about $0.02L_{geom}$ (annual $>
10$ fb$^{-1}$),  that is only 5 times less than that for the standard case  $E=250$~GeV ($\sqrt{s_{\ggam} }\approx 0.4$~TeV).
Recall that in the standard case the spread of effective \ggam \ masses is much wider. Maximal  \ggam \ luminosity
of this HE PLC is more than 25\% from that for the standard case  $E=250$~GeV.
The necessary laser flash energy is $(1.7\div 1.2)A_0$, where $A_0$ is that for standard case   $E=250$~GeV.

4. The discussed values of luminosity show that the number of events  of interest per bunch crossing is much less than 1.  Therefore, signal and background events will be observed separately.

5. The low-energy part of the photon spectrum includes photons from all mechanisms \eqref{processes},
as well as the equivalent photons from $e^-e^-$ scattering and photons from radiation in focusing systems of LC, etc. It is highly dependent on the details of the experimental device.
The corresponding luminosity integral can be quite large \cite{Tel13}.
This part { of spectrum} can be used for more traditional tasks similar to those in the hadron collider (see for example \cite{FCCall}).

\section*{Acknowledgment}

We are thankful to   V.~Serbo and V.~Telnov for discussion and comments.
The work was supported by the program of fundamental scientific
 researches of the SB RAS \# II.15.1., project \# 0314-2019-00 and HARMONIA project under contract   UMO-2015/18/M/ST2/00518 (2016-2020).

\appendix

\section{Some backgrounds and related problems}
\subsection{The case of "bad" initial helicity $\Lambda_C\approx 1$}\label{secbad}
At $\Lambda_C=1$
the  photon spectrum in  the main Compton process is much flatter than in the "good"  case $\Lambda_C = -1 $, see fig.~\ref{x918spectra}. Therefore, in estimates of integrated luminosity \eqref{Lumintdef} one should use lower value $y_{min}=0.6$.  Optimum optical length in
this case is higher than that for the "good" case $ \Lambda_C = -1 $,
in particular, $z_m(x=18,\,\Lambda_C=1)=0.827$ and
$z_m(x=100,\,\Lambda_C=1)=1.07$ for $x=100$.
This requires a laser flash energy that is not much higher than that  for the “good case”, $ A = 1.43A_0 $ for $ x = 18 $ and $ A = 5.4A_0 $ for
$ x = 100$.

The more important difference is the shape of the luminosity spectrum. This spectrum is much flatter than  shown in Fig.~\ref{lum18}.
Its smoothed maximum is shifted to much smaller values of $w$. Here simple one-parametric description of beam collisions \eqref{rhodef}-\eqref{Lggam} become invalid, details of device construction are essential. In this case, the high-energy and low-energy parts of the luminosity spectrum are practically not separated.

With the growth of distance $b$ (fig.~\ref{fig:basschem}) the low-energy part of luminosity disappears,
the residual peak gives much lower integrated luminosity than that at $ \Lambda_C \approx -1 $.

\subsection{The linear polarization of high energy photon}\label{seclinear}

The linear polarization of high energy photon is expressed via
linear polarization of the laser photon $P_{\ell 0}$ by well known ratio $(N/D)$ (see \cite{GKST}),
in which $|N|\le | 2P_{\ell 0}|$ and $D\propto f(x,y)$.
To reach maximal high energy luminosity in the entire spectrum, the denominator $D$ should be large. Therefore, the linear polarization of photon can be only small in the cases with relatively high \ggam luminosity. One cannot hope to study its effects in the considered HE PLC.

\subsection{Collisions of positrons with electrons of  the opposite beam}\label{appa}

Collisions of positrons from killing process with electrons of  the opposite beam result in physical states similar to those produced in \ggam collisions. It will be  the main background for { HE PLC}.\\

{\bf General.}
According to Table~\ref{1bbb}, at  $x=18$ and $z=z_m$ the number of killed photons producing high energy \epe \ pairs is less than 3/4 from the number of operative photons. This ratio decreases at lower $x$. Only one half from these photons produces high energy positrons.  Therefore the luminosity of these collisions\lb ${\cal L}_b(\epe)\lesssim (1.5\div 2.5){\cal L}(\ggam)$.

$\lozenge$  The use the optical length $z_{0.9}$ instead of the optimal one $z_m$  reduces  number of positrons and ${\cal L}_b(\epe)$   by half or even more  with  a small change  in  $L_{\ggam}$.\\

{\bf Distribution of positrons etc.}  The details of luminosity distribution of these \epe collisions differ
 strongly from those for \ggam collisions of main interest. To see these details we discuss  the energy distribution of positrons, produced in collision of the laser photon with the high energy photon having energy $\omega=Ey$ and polarization $\lambda$. We use notations \eqref{killproc} and denote  the positron energy $E_+=y_+\omega$. First of all,  recall the kinematic variables:
\bes\label{kinemkill}\be
w^2=sy\ge 4,\quad v=\sqrt{1-4/w^2}, \quad \fr{1+v}{2}\ge y_+\ge \fr{1-v}{2},
\label{kinemkill1}
\ee
therefore,
\beg
y_+\le 0.77\to E_+\le 0.693E \;\; at\;\; x=9,\\
y_+\le 0.888\to E_+\le 0.841 \;\; at\;\; x=18\,.\label{killkinem2}
\eeg\ees
As a result,  the rapidity of system produced in these \epe collision is
\be
\eta_{\epe}(x=9)\ge 0.153 ,\qquad
\eta_{\epe}(x=18)\ge 0.08. \label{rapidepe}
\ee
These values don't intersect with  the possible rapidity interval for \ggam \ system
\eqref{rapidggam}, \eqref{taudef}. Therefore,  the \epe and \ggam \ events are  clearly distinguishable
in the case  of observation of all reaction products.

In general, a more detailed description is desirable.  The energy distribution of positrons produced in the $\gamma_o\gamma$ collision is
\beg
\fr{dn_+(y_+;w^2,\lambda\lambda_o)}{dy_+}
= \fr{(u-2)(1+c\lambda_o\lambda)+s^2}
{\Phi_{\ggam}(w^2,\lambda_o\lambda)};\\
u=\fr{1}{y_+(1-y_+)},\quad c=2\fr{u}{w^2}-1,\quad s^2=1-c^2\,.
\label{posdistrggam}\eeg
For the  highest positron energy  $c=1$, and we have $dn=0$  (at\lb $\lambda_o\lambda=-1$). This
equality $dn=0$  corresponds to the fact that the angular momentum conservation forbids production of positrons (or electrons) in the forward direction. It means that the physical flux of positrons is limited even stronger than that given by Eq.~\eqref{kinemkill}.

Except mentioned endpoints, distribution \eqref{posdistrggam} changes weakly in the whole interval of $y_+$ variation. Therefore, the \epe luminosity is widely distributed over  entire range of its possible variation.  As a result,  the differential luminosity $dL/dw_{\epe}\ll dL/dw_{\ggam}^{peak}$.\\

{\bf Difference in the produced sets of particles.}
 In addition to the difference in the distribution of luminosity of $\ggam$ and \epe collisions we enumerate some characteristic differences in the produced systems for identical energy.

 1. With the growth of energy all cross sections in \epe mode  decrease as $1/s$. In the \ggam mode cross sections of many processes don't decrease.

 2. In the \epe mode most of processes are annihilation  ones (via $\gamma$ or $Z$ intermediate states). Products of reaction in such processes have wide angular distribution. In the \ggam mode the essential part of produced particles moves along the collision axis, small transverse momenta are favorable.

\section{Some physical problems for HE PLC}\label{secphys}

We expect that LHC and $e^+e^-$ LC will give us many new results.  Certainly, HE PLC complements these results and improve precision of some fundamental parameters. If some new particles would be discovered, HE PLC
 will also allow  to improve the precision of their parameters. We will list here only  problems  for which the HE PLC can provide fundamentally new information that cannot be reduced to a simple refinement of the results obtained at the LHC and \epe \ LC (see for example \cite{FCCall}), \cite{TESLA,ILCmod,CLICmod}).

\bu {\bf Beyond Standard Model.}
In the extended Higgs sector one can realizes scenario, in which the observed Higgs boson $h$ is the SM-like (aligned) particle, while model contains  others scalars which interact strongly. It was discussed earlier (for a minimal SM with one Higgs field) that physics of such strongly interacting Higgs sector can be similar to a low-energy pion physics. Such  a system may have resonances like $\sigma$, $\rho$, $f$ with spin 0, 1 and 2 (by estimates, with mass $M\lesssim 1-2$~TeV). High monochromaticity of HE PLC allows to observe  these  resonance states with spin 0 or 2  in \ggam \ mode.

In the same manner one can observe excited electrons with spin 1/2 or 3/2 in \egam \ mode.

\bu {\bf Gauge boson physics.}

The SM electroweak theory is checked now  at the tree level for simplest processes and at the 1-loop level for $Z$-peak. The opportunity to test effects of this model for more complex processes and for loop corrections beyond $Z$-peak looks very  significant . HE PLC provide us  with a unique opportunity to  study these problems.

$\triangledown$ Processes \ $\gamma\gamma\to WW$, $e\gamma\to \nu W$ have huge cross sections, at large $s$ we have
\bes\label{gauge2}\bea
\sigma(\gamma\gamma\to WW)\equiv \sigma_V=8\pi\alpha^2/M_W^2\approx 86~pb\,,\label{ggwwcr}\\
\sigma(e\gamma\to \nu W)\approx\left(\fr{1}{2}-\lambda_e\right) \sigma_V\,,\label{egWnucr}\\
\sigma(\gamma\gamma\to ZZ)\sim \alpha^2\sigma_V\sim (1\div 10)~fb\,.
\eea\ees
 Therefore the measurement  of these processes allows  us  to
test the  detailed structure of the electroweak theory
with  an  accuracy about 0.1\%  (1-loop and partially 2-loop). To describe properly these results with such  an accuracy, one should solve the fundamental QFT problem  on the  perturbation theory with unstable particles. With such precision, the sensitivity to possible anomalous interactions (operators of higher dimension)  --  that is, to the signals of BSM physics --  will be  enhanced \cite{GKS_W}.  In this connection, we may mention  the $\ggam\to ZZ$ process, for which only  a rough estimate of the cross section is presented. This is the first well-measurable process with variable energy, obliged by only loop contributions.

$\triangledown$ Processes with {\bf multiple production of gauge bosons} at HE PLC have relatively large cross sections, being sensitive to the details
of gauge boson interactions (which cannot be seen in another way) and possible anomalous    interactions.         Nothing similar can be offered by other colliders.  The relatively large values of the cross sections are due to the contribution of the diagrams with the exchange of vector bosons in the $t-$ channel, which does not decrease with increasing $s$ \cite{GIS_3}.
Moreover, at a sufficient distance from the reaction threshold,  these cross sections grow logarithmically with factors $L\approx \ln (s/m_e^2)$ for  photon exchange  or  $\ell\approx \ln(s/M_W^2)$ for  $W(Z)$ exchange :
\beg
\sigma(\egam\to eWW)\sim\alpha\sigma_W L\,,\quad \sigma(\egam\to eWWZ)\sim\alpha^2\sigma_W L\ell\,,\\
\sigma(\egam\to\nu WW)\sim\alpha\sigma_W \ell\,, \quad\sigma(\ggam\to ZWW)\sim\alpha\sigma_W \ell\,,\\
\sigma(\egam\to\nu WWW)\sim \fr{\sigma(\ggam\to WWZZ)}{\sin^2\theta_W}\sim\alpha^2\sigma_W \ell^2\,,
\\ \sigma(\ggam\to WWWW)\sim
\alpha^2\sigma_W \ell^2\,.
\eeg
\begin{figure}[hbt]
\includegraphics[width=0.48\textwidth,height=0.25\textheight]{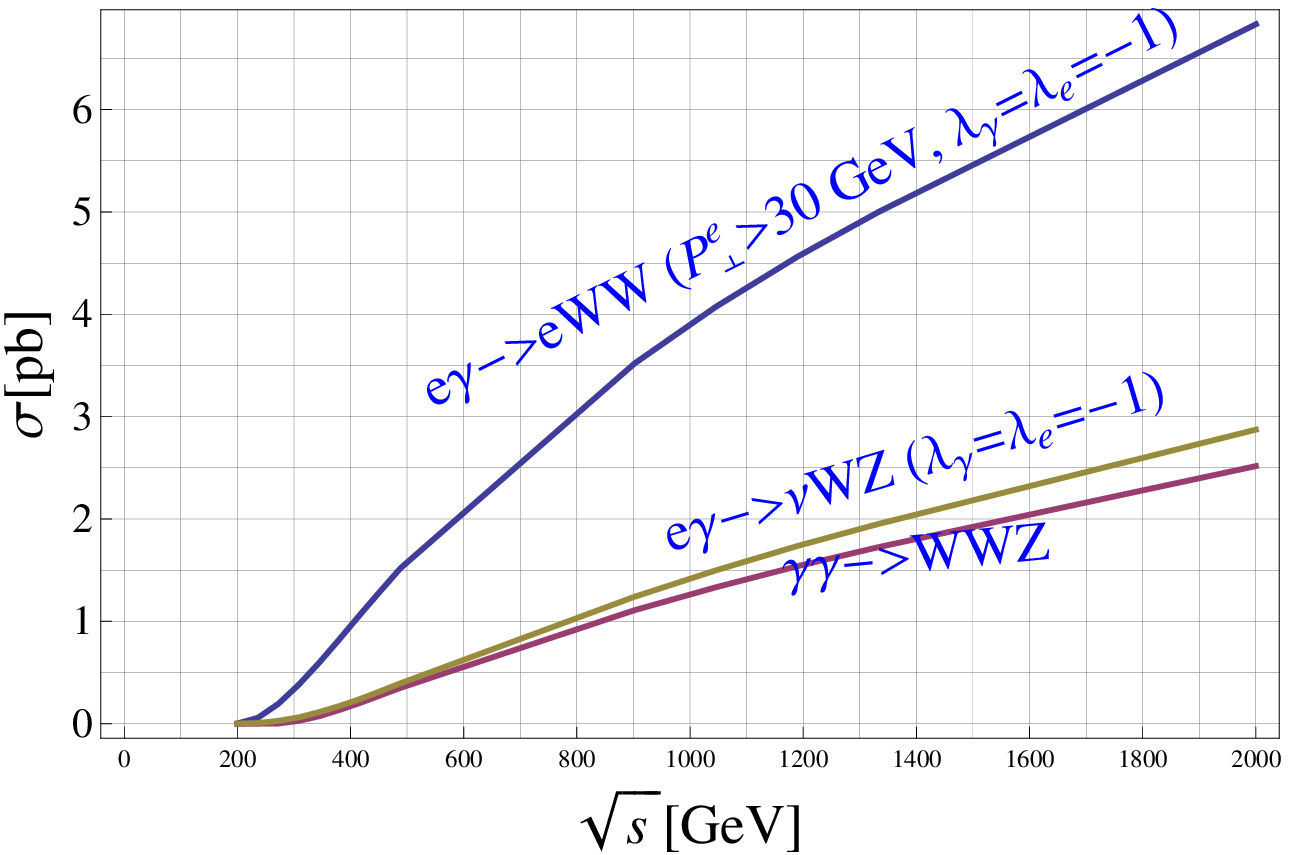}\hspace{5mm}
\includegraphics[width=0.48\textwidth,height=0.25\textheight]{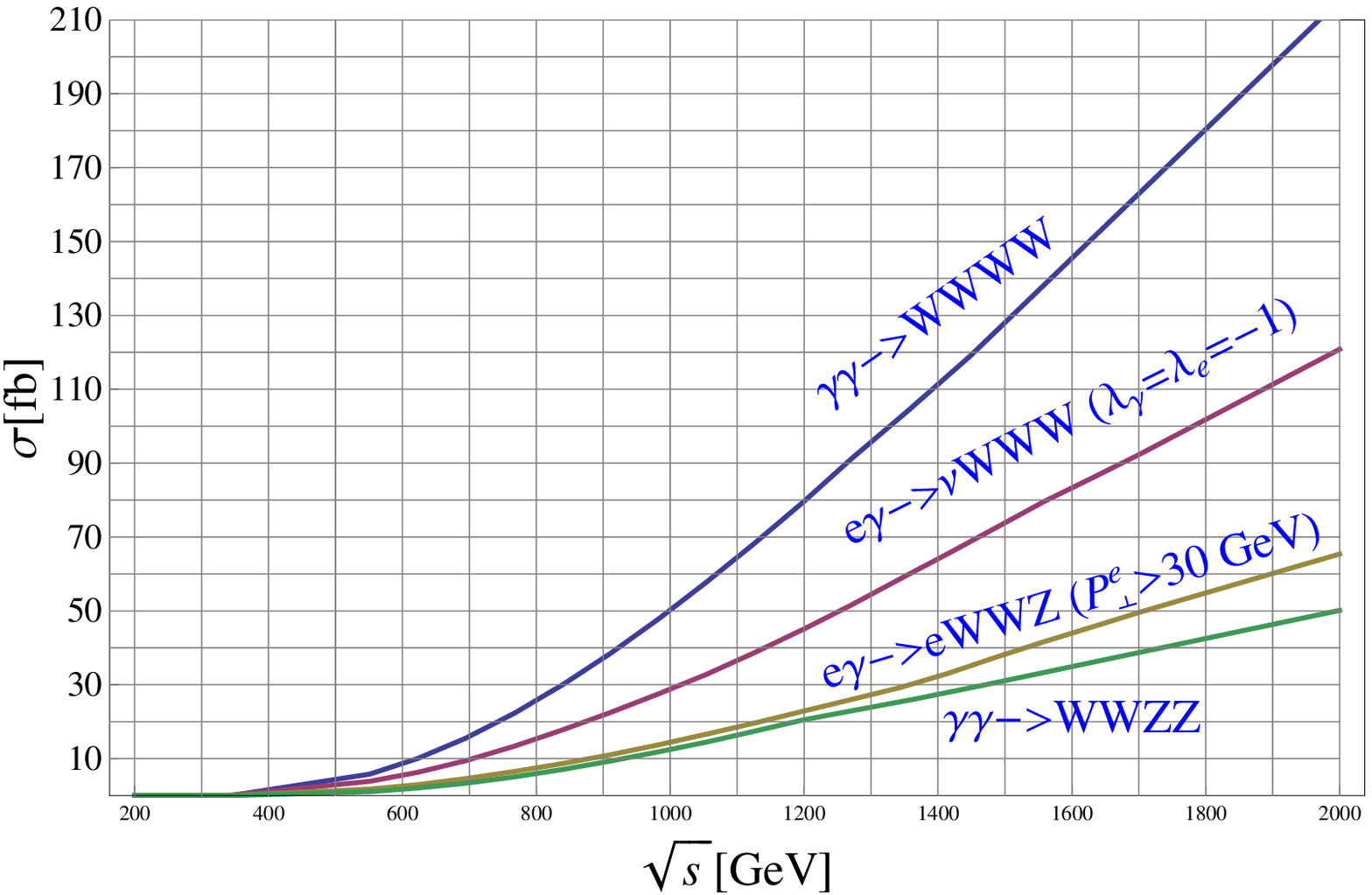}
\caption{\it Cross sections  for multiple production of gauge bosons, calculated with CalcHEP.}
  \label{PROCESS1}
\end{figure}
These cross sections are shown in Fig.~\ref{PROCESS1}. For processes $\egam\to eWW$, $\egam\to eWWZ$ we present cross sections for well observable transverse momenta of electrons ($p_{\bot e}>30$~GeV). Studying the dependence of these cross sections on the electron transverse momentum will allow to measure the electromagnetic form factor $W$ in the  processes\lb $\gamma\gamma^*\to WW$, $\to WWZ$ and separate the contribution of processes $Z^*\gamma\to WW$, $Z^*\gamma\to WWZ$ (at $p_{\bot e}>30$~GeV).
The study of  process $\egam\to \nu WZ$ allows to extract first information about subprocess $W^*\gamma\to WZ$.

\bu {\bf Hadron physics and QCD.}
Our understanding of hadron physics is twofold. We believe that we understand basic theory -- QCD with its asymptotic freedom. However, the  results of calculations in QCD can be applied to the description of data only with the aid some phenomenological assumptions (often verified by long practice). It results in badly controlled uncertainties in the description of data.

$\triangledown$ PLC is to some extent the hadronic machine with more pure initial state than LHC. Therefore, PLC  can be used also  for detailed study of high energy QCD processes like diffraction, total cross sections, odderon, etc. The results of such experiments can be confronted to  theory with much lower uncertainty than the corresponding ones at the LHC.

$\triangledown$ The study  of  the photon structure function $W_\gamma$  (in \egam \ mode) provides a  unique test of QCD. This function can be written as the sum of point-like $W_\gamma^{pl}$ and hadronic $W_\gamma^h$ contributions.  The hadronic part is similar to that for proton and it describes with similar phenomenology. In contrast, the point-like contribution $W_\gamma^{pl}$ is described  without phenomenological  parameters \cite{Wit}. The ratio  $W_\gamma^h/W_\gamma^{pl}$ decreases with $Q^2$ roughly as $(\ln Q^2)^{-1/3}$. In the range of parameters accessible today the hadronic contribution dominates. With the growth of  $Q^2$ at \egam \ HE PLC the point-like part becomes dominant.

\end{document}